\documentclass[11pt,a4paper]{article}

\usepackage{latexsym}
\usepackage{amssymb}
\usepackage{amsmath}
\usepackage[authoryear]{natbib}

\newif\ifpdf
\ifx\pdfoutput\undefined
   \pdffalse
\else
   \pdfoutput=1
   \pdftrue
\fi

\ifpdf
    \usepackage{color}
    \definecolor{myred}{rgb}{0.5,0,0}
    \definecolor{myblue}{rgb}{0,0,0.75}
    \definecolor{mygreen}{rgb}{0,0.5,0}
   \usepackage[pdftex,
               pdftitle={Expected Shortfall and Beyond},
               pdfsubject={Special coherent risk measures},
               pdfkeywords={Expected Shortfall; Value-at-Risk; Spectral
                            Measure; coherence; risk contribution; marginal impact},
               pdfauthor={Dirk Tasche},
               pdfstartview=FitBH,
               breaklinks=true,
               colorlinks=true,
               citecolor=myred,
               linkcolor=myblue,
               urlcolor=mygreen]{hyperref}
\else
   \usepackage{hyperref}
\fi

\newtheorem{theorem}{Theorem}[section]

\newtheorem{remark}[theorem]{Remark}
\newtheorem{example}[theorem]{Example}
\newtheorem{proposition}[theorem]{Proposition}
\newtheorem{definition}[theorem]{Definition}
\newtheorem{corollary}[theorem]{Corollary}

\numberwithin{equation}{section}
\title{Expected Shortfall and Beyond}
\author{%
Dirk Tasche\thanks{
Deutsche Bundesbank,
Postfach 10 06 02,
60006 Frankfurt\ a.\ M.,
Germany; e-mail: tasche@ma.tum.de\newline
The contents of this paper do not necessarily 
reflect opinions shared by Deutsche Bundesbank.}
}

\date{October 20, 2002}

\begin{document}
\maketitle

\begin{abstract}
Financial institutions have to allocate so-called \emph{economic capital}
in order to guarantee solvency to their clients and counterparties.
Mathematically speaking, any methodology of allocating capital is
a \emph{risk measure}, i.e.\ a function mapping random variables to the real numbers.
Nowadays \emph{value-at-risk}, which is defined as a fixed level quantile of
the random variable under consideration, is the most popular risk measure.
Unfortunately, it fails to reward diversification, as it is not \emph{subadditive}.

In the search for a suitable alternative to value-at-risk,
\emph{Expected Shortfall} (or \emph{conditional value-at-risk} or \emph{tail value-at-risk})
has been characterized as the smallest \emph{coherent} and \emph{law invariant} risk
measure to dominate value-at-risk. We discuss these and some other properties
of Expected Shortfall as well as its generalization to a class of coherent risk
measures
which can incorporate higher moment effects. Moreover, we suggest a general method
on
how to attribute Expected Shortfall \emph{risk contributions} to portfolio components.\\[1ex]
\textbf{JEL classification} D81, C13.\\[1ex]
\textbf{Keywords} Expected Shortfall, Value-at-Risk, Spectral Risk Measure, coherence, risk
  contribution.
\end{abstract}

\section{Introduction}
\label{sec:intro}

At the latest in 1999, when the article \citet{ADEH99} appeared, it became
clear that value-at-risk (see Definition \ref{def:1} below) cannot be
considered a sound methodology for allocating economic capital in financial
institutions. However, even if in \citet{ADEH99} recommendations were given for
the properties sound risk measures should satisfy, only recently Expected
Shortfall
(Definition \ref{de:10} below) was suggested as practicable and sound
alternative to value-at-risk. Nevertheless, there are still a lot of useful
properties of Expected Shortfall and its generalizations which cannot be found
in printed sources so far. 

With the paper at hand, we try to make up for this omission. We will
recapitulate in section~\ref{sec:VaR} what makes value-at-risk a seductive
measure of risk and what are the main criticisms against it. In particular, we
will see a new example (Example \ref{ex:1}) for its lacking subadditivity and give
a new interpretation (Remark \ref{rm:10}) why this is an important point.

We will then introduce in section~\ref{sec:spec} Expected Shortfall as a
convincing alternative to value-at-risk. We will summarize some of its more
important properties. These properties are shared by all the representatives
of the class of spectral risk measures that were introduced in
\citet{acerbi} (cf.\ Remark \ref{rm:12} below). Generalizing a
result from \citet{kusuoka}, we show that all the elements of this class can be
represented as certain averages of values-at-risk at different levels (Theorem
\ref{th:1}). This representation allows the easy creation of risk measures
which enjoy the useful properties of Expected Shortfall and incorporate
other desirable features like moment effects.

When a risk measure for a portfolio has been chosen the question arises how to attribute risk
contributions to subportfolios. This is of interest for a risk diagnostics of
the portfolio \citep[see][]{L96} or for performance analysis. In
section~\ref{sec:contr}, we present a suggestion of how to do this in case of
spectral risk measures (Definition \ref{de:12} and Proposition
\ref{pr:14}). Finally,
we show for the Expected Shortfall that these contributions can be interpreted
as conditional expectations given a worst case scenario (Proposition \ref{pr:2}).

\section{Value-at-Risk: lacking subadditivity}
\label{sec:VaR}

Consider a random variable $X$ which might be seen as the random profit and
loss of an investment by a fixed time horizon. Positive values of $X$ are
regarded as profits, negative values as losses. The value-at-risk (VaR) of $X$
at level $\alpha$ is the absolute value of the worst loss not to be exceeded
with a probability of at least $\alpha$. The following couple of definitions
gives a formal description of this quantity.
\begin{definition}[Quantile, value-at-risk]
  \label{def:1}
Let $\alpha \in (0,1]$ be fixed and $X$ be a real random variable on a
probability
space $(\Omega, \mathcal{F}, \mathrm{P})$. Define $\inf \emptyset =
\infty$. We then call
\begin{subequations}
\begin{equation}
  \label{eq:31}
  q_\alpha(X) \ =\ \inf\bigl\{ x\in \mathbb{R}:\,\mathrm{P}[X \le x] \ge
  \alpha \bigr\}
\end{equation}
the \emph{$\alpha$-quantile} of $X$. We call
\begin{equation}
  \label{eq:32}
  \mathrm{VaR}_\alpha(X) \ = \ q_\alpha(- X)
\end{equation}
the \emph{value-at-risk (VaR)} at (confidence) level $\alpha$ of $X$.
\end{subequations}
\end{definition}
Usually, values of $\alpha$ close to 1 are of interest. Since by definition
$\mathrm{P}\bigl[X + \mathrm{VaR}_\alpha(X) \ge 0\bigr] \ge \alpha$,
$\mathrm{VaR}_\alpha(X)$ can be interpreted as the minimal amount of capital
to be put back by the investor in order to preserve her solvency with
a probability of least $\alpha$.

Below, we will compare $\mathrm{VaR}$ to other methods for attributing capital
to random variables (sometimes in insurance contexts also called risks). A
positive capital attribution means that the risk under consideration
\emph{requires} capital whereas a negative capital attribution indicates that
capital may be released. From an economical point of view, it makes sense to
allow for risks which require a positively infinite amount of capital. A risk
with capital requirement $\infty$ must not be accepted by the investor. The
interpretation of a risk with capital requirement $- \infty$ is much less
straightforward.
Would this imply that such a risk can serve as a collateral for any risk with
finite capital requirement? However, this case does not appear very likely and
is therefore excluded  from the following definition of risk measures.
\begin{definition}[Risk measure]
  \label{def:2}
Let $(\Omega, \mathcal{F}, \mathrm{P})$ be a probability space and $V$ be a
non-empty set of $\mathcal{F}$-measurable real-valued random variables. Then
any mapping $\rho : \, V\to \mathbb{R}\cup \{\infty\}$ is called a \emph{risk measure}.
\end{definition}
$\mathrm{VaR}$, as a risk measure in the sense of Definition \ref{def:2}, enjoys
most of the properties that are considered useful in the literature
\citep{ADEH99, kusuoka}. 
\begin{proposition}[Properties of value-at-risk]
  \label{pr:10}
Let $\alpha \in (0,1]$ be fixed and 
$(\Omega, \mathcal{F}, \mathrm{P})$ be a probability
space. Consider the risk measure $\rho$ on the set $V$ of all the 
$\mathcal{F}$-measurable real-valued random variables which is given by
\begin{equation}
  \label{eq:33}
  \rho(X)\ = \ \mathrm{VaR}_\alpha(X), \quad X \in V.
\end{equation}
Then $\rho$ has the following properties:
\begin{enumerate}
\item[(1)] Monotonicity: $X, Y \in V$, $X \le Y$ \ $\Rightarrow$\ $\rho(X) \ge
  \rho(Y)$.
\item[(2)] Positive homogeneity: $X\in V, h > 0, h\,X\in V$ \ $\Rightarrow$\
  $\rho(h\,X) = h\,\rho(X)$.
\item[(3)] Translation invariance: $X\in V, a \in\mathbb{R}, X + a \in V$\
  $\Rightarrow$\ $\rho(X+a) = \rho(X) -a$.
\item[(4)] Law invariance: $X, Y \in V$, $\mathrm{P}[X \le t] = \mathrm{P}[Y
  \le t]$ for all $t\in \mathbb{R}$ \\
  $\Rightarrow$\ $\rho(X) = \rho(Y)$.
\item[(5)] Comonotonic additivity: $f, g$ non-decreasing, $Z$ real random
  variable
on $(\Omega, \mathcal{F}, \mathrm{P})$ such that $f\circ Z, g\circ Z \in V$
\\ $\Rightarrow$\ $\rho(f\circ Z + g\circ Z) = \rho(f\circ Z) + \rho(g\circ Z)$.
\end{enumerate}
\end{proposition}
\noindent \textbf{Proof.} (1) until (4) are obvious. For (5), see
e.g.~\cite{De}.\hfill$\Box$

Note that $\mathrm{VaR}_\alpha$ is law invariant in a very strong sense: the
distributions of $X$ and $Y$ need not be identical in order to imply
$\mathrm{VaR}_\alpha(X) = \mathrm{VaR}_\alpha(Y)$. A certain local identity of
the distributions suffices for this implication. In particular, random
variables $X$ with light tail probabilities and $Y$ with heavy tail
probabilities \citep[see e.g.][]{EKM} may have the same
$\mathrm{VaR}_\alpha$. This point is one main criticism against $\mathrm{VaR}$
as a risk measure.

One important property is missing in the enumeration of Proposition
\ref{pr:10}: the \emph{subadditivity}, i.e.
\begin{equation}
  \label{eq:34}
X, Y \in V, X + Y \in V \ \Rightarrow\   \rho(X+Y) \le \rho(X) + \rho(Y).
\end{equation}
It is well-known that $\mathrm{VaR}$ is \emph{not} in general subadditive.
Here we present a counterexample with continuous and even independent random
variables.
\begin{example}[Lacking subadditivity of VaR]
  \label{ex:1}
  \begin{subequations}
Let $X_1, X_2$ Pareto distributed with values in  $(-\infty, 1)$ and
independent. The
joint distribution of $(X,Y)$ is specified by
\begin{equation}
\label{eq:35}
\mathrm{P}[X_1 \le x_1, X_2 \le x_2]\ =\  (2-x_1)^{-1} (2-x_2)^{-1},\quad x_1,
x_2 < 1. 
\end{equation}
This implies
\begin{alignat}{2}
\label{eq:36}
\mathrm{VaR}_\alpha(X_i) &\ = \ (1-\alpha)^{-1} -2,&\qquad & i= 1,2,\\
\mathrm{P}[X_1 + X_2 \le x] & \ =\ \frac{2}{4-x} + \frac{2\,\log(3-x)}{(4-x)^2},&\qquad & x < 2.\notag  
\end{alignat}
By (\ref{eq:36}), we have $\mathrm{VaR}_\alpha(X_1) + \mathrm{VaR}_\alpha(X_2)
< \mathrm{VaR}_\alpha(X_1+X_2)$ for all $\alpha \in (0,1)$ because
\begin{equation}
  \label{eq:37}
\begin{split}
\mathrm{P}\bigl[- (X_1 + X_2) \le 2\,\mathrm{VaR}_\alpha(X_1)\bigr]& \ =\ 
\textstyle \alpha - \frac{1-\alpha}2\,\log\frac{1+\alpha}{1-\alpha} \\
&\ < \ \alpha.
\end{split}
\end{equation}
In particular, for $\alpha = 0,99$ we have
\begin{equation*}
\mathrm{VaR}_\alpha(X_1) = \mathrm{VaR}_\alpha(X_2) = 98,\quad \mathrm{VaR}_\alpha(X_1+X_2) \approx 203,2.
\end{equation*}  
\end{subequations}
\end{example}
The lacking subadditivity of $\mathrm{VaR}$ is criticized because under
certain circumstances it might be an incentive to split up a large firm into
two smaller firms. Another interpretation (Remark \ref{rm:10}) follows from the following result.
\begin{proposition}
  \label{pr:11}
Let $X, Y$ be real, linearly independent random variables and $\rho$ be a
real-valued
risk measure on the positive cone $C$ spanned by $X$ and $Y$, i.e. $C = \bigl\{ u\,X +
v\,Y :\,u,v > 0\bigr\}$. Assume that $\rho$ is positively homogeneous in the
sense of Proposition \ref{pr:10}~(2) and that the function $\rho(u,v) = \rho( u\,X +
v\,Y)$, $u,v > 0$ is differentiable in $(u,v)$. Then we have 
\begin{subequations}
  \begin{equation}
    \label{eq:38}
    \rho(U_1 + U_2) \ \le \ \rho(U_1) + \rho(U_2), \quad U_1, U_2 \in C,
  \end{equation}
if and only if
\begin{equation}
  \label{eq:39}
  \rho_{U_1}(U_1 + U_2) \le \rho(U_1),\ \rho_{U_2}(U_1 + U_2) \le \rho(U_2),
\quad \quad U_1, U_2 \in C,
\end{equation}
with 
\begin{equation*}
\rho_{U_i}(U_1 + U_2)  = u_i\,\frac{\partial \rho}{\partial u}\bigl(u_1 + u_2,
v_1+ v_2\bigr) +  v_i\,\frac{\partial \rho}{\partial v}\bigl(u_1 + u_2,
v_1+ v_2\bigr)
\end{equation*}
when $U_i = u_i\,X + v_i\,Y$, $i =1, 2$ (note that by linear independence this
representation is unique).
\end{subequations}
\end{proposition}
\begin{remark}
  \label{rm:10}
By Euler's relation (see (\ref{euler})), the terms $\rho_{U_1}(U_1 + U_2)$
and $\rho_{U_2}(U_1 + U_2)$ from (\ref{eq:39}) sum up to $\rho(U_1 +
U_2)$. Hence it appears quite natural to regard them as the risk (or capital)
contributions
of $U_1$ and $U_2$ respectively to the total capital $\rho(U_1 + U_2)$ which
is required by $U_1 + U_2$.
Indeed, it can be argued that there is no other way to arrive at a reasonable
notion of capital contribution than by partial derivatives \citep[cf.][]{Denault, T99}. 
Moreover, $\mathrm{VaR}$ and the risk measure $\mathrm{ES}$ to be defined below
(Definition \ref{de:11}) satisfy the conditions of Proposition \ref{pr:11}
under quite general assumptions on the joint distribution of $(X,Y)$ \citep[cf.][]{T00}.

With this interpretation of $\rho_{U_i}(U_1 + U_2)$, $i = 1,2$, the meaning of
(\ref{eq:39}) is as follows: the manager who is responsible for subportfolio
$U_1$ will never be damaged by diversification in the portfolio of the firm
because her capital contribution will never be greater than the capital
requirement in the case of $U_1$ considered as a stand-alone portfolio.
\end{remark}
\noindent \textbf{Proof of Proposition \ref{pr:11}.} We show first that (\ref{eq:38})
implies $\rho_{U_1}(U_1 + U_2) \le \rho(U_1)$. Fix $U_i = u_i\,X + v_i\,Y$, $i
=1, 2$, and note that $\rho(U_1) = \rho(u_1, v_1)$ and $\rho(U_1+U_2) =
\rho(u_1+u_2, v_1+v_2)$.
Define the function $f : (-1, \infty) \to \mathbb{R}$ by
\begin{subequations}
  \begin{align}
\begin{split}
    f(t) & = \rho(u_1+u_2, v_1+v_2) + t\,\rho(u_1, v_1) \\
& \qquad -\,\rho\bigl((1+t)\,u_1+u_2, (1+t)\,v_1+v_2\bigr).
\end{split}
\intertext{Then}
\begin{split}
f'(t) & = \rho(u_1, v_1) 
 - u_1\,\frac{\partial \rho}{\partial u}\bigl((1+t)\,u_1 + u_2,
(1+t)\,v_1+ v_2\bigr)\\
& \qquad  -\,  v_1\,\frac{\partial \rho}{\partial v}\bigl((1+t)\,u_1 + u_2,
(1+t)\,v_1+ v_2\bigr)
\end{split}
\intertext{and in particular}
\begin{split}
\label{eq:40}
  f(0) & = 0,\\
f'(0) & = \rho(u_1, v_1)
 - u_i\,\frac{\partial \rho}{\partial u}\bigl(u_1 + u_2,
v_1+ v_2\bigr) 
 +  v_i\,\frac{\partial \rho}{\partial v}\bigl(u_1 + u_2,
v_1+ v_2\bigr)\\
& = \rho(U_1) - \rho_{U_1}(U_1+U_2).
\end{split}
  \end{align}
(\ref{eq:38}) implies for $t >0$ that $f(t) \ge 0$. But, by (\ref{eq:40}),
this is a contradiction to the assumption $\rho(U_1) - \rho_{U_1}(U_1+U_2) =
f'(0) < 0$. This implies (\ref{eq:39}).

Let us now consider the proof of the implication (\ref{eq:39}) $\Rightarrow$
(\ref{eq:38}).
This is easy since by Euler's relation and (\ref{eq:39})
\begin{align}\label{euler}
  \begin{split}
 \rho(U_1 + U_2) & =   \rho(u_1+u_2, v_1+v_2)\\
& =  (u_1 + u_2)\,\frac{\partial \rho}{\partial u}\bigl(u_1 + u_2,
v_1+ v_2\bigr)\\
& \qquad 
 - \, (v_1+v_2)\,\frac{\partial \rho}{\partial v}\bigl(u_1 + u_2,
v_1+ v_2\bigr)\\
&\le \rho(U_1) +\rho(U_2).
  \end{split}
\end{align}
This completes the proof of Proposition \ref{pr:11}. \hfill $\Box$
\end{subequations}

\section{Spectral risk measures}
\label{sec:spec}

The weak points of VaR as a risk measure are well-known for some time
\cite[cf.][]{ADEH99}. Nowadays, there is a certain consensus on the properties
a reasonable risk measure should satisfy \citetext{\citealp{ADEH97, ADEH99, Delb98};
but see also \citealp{FSch}, for a relaxation}: 
it should be coherent in the sense of the following
definition.
\begin{definition}[Coherent risk measure]
  \label{de:10}
A risk measure $\rho : V \to \mathbb{R}\cup\{\infty\}$ in the sense of
Definition \ref{def:2} is called \emph{coherent} if it is monotonous,
positively homogeneous, translation invariant, and subadditive (see
Proposition \ref{pr:10} (1), (2), (3), and Eq.~(\ref{eq:34})).
\end{definition}
In order to preserve the desirable connection between the level of
$\mathrm{VaR}_\alpha$ and the investor's probability of solvency, it would be
nice to have a smallest coherent risk measure to dominate
$\mathrm{VaR}_\alpha$. As was shown in \cite{Delb98}, such a smallest coherent
majorant to $\mathrm{VaR}_\alpha$ does not exist. Nevertheless, in
\cite{Delb98} was also shown there is a smallest coherent and 
\emph{law invariant} 
(see Proposition \ref{pr:10} (4))
risk measure\footnote{%
The term ``law invariance'' was introduced in \cite{kusuoka}. A rough
interpretation of law invariance might be ``can be estimated from statistical observations only''. 
Anyway, as VaR is law invariant it seems natural to look
for its smallest coherent \emph{and} law invariant majorant.
See eq.\ (1) in \citet{acerbi} for an example of a risk measure which is not law invariant
in sense of Proposition \ref{pr:10}~(4).
}%
that dominates $\mathrm{VaR}_\alpha$. The representation of this measure in
\cite{Delb98} was not explicit in the general case. However, it became clear
that for continuous random variables $X$, it coincides with $\mathrm{E}\bigl[
- X\,|\,- X \le \mathrm{VaR}_\alpha(X)\bigr]$, the so-called
\emph{tail value-at-risk}. Note that tail value-at-risk, in general, is not
subadditive \citep[see e.g.][]{AT01}.

Denote -- as usual -- by $\mathbf{1}_A = \mathbf{1}_A(a)$ the \emph{indicator
  function} of the set $A$, i.e. $\mathbf{1}_A(a) = 0$ if $a \notin A$ and 
$\mathbf{1}_A(a) =1$ if $a \in A$.
\begin{definition}
  \label{de:11}
Let $\alpha \in (0,1)$ be fixed and $X$ be a real random variable on a
probability space $(\Omega, \mathcal{F}, \mathrm{P})$ with
$\mathrm{E}\bigl[\max(0, -X)\bigr] <\infty$. Define $q_\alpha(- X)$
as in Definition \ref{def:1}. We then call 
\begin{multline}
  \label{eq:41}
  \mathrm{ES}_\alpha(X) = - (1-\alpha)^{-1} \Big( \mathrm{E}\bigl[
  X\,\mathbf{1}_{\{ - X \ge q_\alpha(- X)\}}\bigr] \\
+ q_\alpha(- X)\,
\Big\{ \alpha - \mathrm{P}\bigl[ - X < q_\alpha(- X)\bigr]\Big\}\Big)
\end{multline}
\emph{Expected Shortfall (ES)} at level $\alpha$ of $X$.
\end{definition}
It turned out \citep{kusuoka, AT01} that ES from Definition \ref{de:11} is
just the smallest coherent and law invariant majorant of $\mathrm{VaR}_\alpha$
which had been already mentioned in \cite{Delb98}. The term ES stems from
\cite{ANS}
where a further proof of the coherence of ES was given. Independently, ES was
introduced in \cite{RU01} under the notion \emph{Conditional value-at-risk
  (CVaR)}. The properties of ES are discussed in detail in \citet{AT01} and \citet{RU01}.

The following result \citep{AT01, Pf00} is important for the calculation of
VaR and ES, and, by the way, enlightens the relationship between the notion of
ES and the \emph{quantile regression} which was introduced in \cite{KB}. ES is
just the optimal value in an optimization problem where $- \mathrm{VaR}$ is
the optimizing argument.
\begin{proposition}
  \label{pr:12}
For $\mathrm{ES}_\alpha$ as given in Definition \ref{de:11} and $q_\alpha,
  q_{1-\alpha}$
as given in Definition \ref{def:1}, we have 
\begin{subequations}
  \begin{equation}
\mathrm{ES}_\alpha(X)  = \min_{s\in\mathbb{R}}  - (1-\alpha)^{-1} \Big( \mathrm{E}\bigl[
  X\,\mathbf{1}_{\{ - X \ge s\}}\bigr]  + s \Big\{\alpha - \mathrm{P}\bigl[ -
  X < s\bigr]\Big\}\Big)
\end{equation}
and
\begin{multline}
\label{eq:42}
\bigl[q_\alpha(- X), - q_{1-\alpha}(X)\bigr]  = \arg \min_{s\in\mathbb{R}}  - (1-\alpha)^{-1} \Big( \mathrm{E}\bigl[
  X\,\mathbf{1}_{\{ - X \ge s\}}\bigr]\\
 + s \Big\{\alpha - \mathrm{P}\bigl[ -
  X < s\bigr]\Big\}\Big), 
\end{multline}
\end{subequations}
whenever $X$ is a real random variable with $\mathrm{E}\bigl[\max(0, -X)\bigr] <\infty$.
\end{proposition}
\noindent\textbf{Proof.} Proposition 4.2 in \cite{AT01}.\hfill$\Box$

Note that the interval in (\ref{eq:42}) is never empty and that
$\mathrm{VaR}_\alpha(X) = q_\alpha(- X)$ by definition. Let us now have a look
on another useful representation of ES.
\begin{proposition}
  \label{pr:13}
For $\mathrm{ES}_\alpha$ as given in Definition \ref{de:11} and
$\mathrm{VaR}_\alpha$ as given in Definition \ref{def:1}, we have
\begin{equation}
  \label{eq:43}
\mathrm{ES}_\alpha(X)  =  (1-\alpha)^{-1} \int_\alpha^1 \mathrm{VaR}_u(X)\, d u, 
\end{equation}
whenever $X$ is a real random variable with $\mathrm{E}\bigl[\max(0, -X)\bigr] <\infty$.
\end{proposition}
\noindent\textbf{Proof.} Proposition 3.2 in \cite{AT01}.\hfill$\Box$

In combination with Proposition \ref{pr:10}, Proposition \ref{pr:13} implies
that
ES is a law invariant and comonotonic additive risk measure. The comonotonic
additivity of a risk measure becomes particularly interesting when it occurs
at the same time as subadditivity. 
\begin{remark}
  \label{rm:11}
Fix $\alpha \in (0,1)$ and consider integrable random variables $X$ and
$Y$. Assume that we do not know the joint distribution of $X$ and $Y$. Then,
from subadditivity, we see that $\mathrm{ES}_\alpha(X) +
\mathrm{ES}_\alpha(Y)$ is an upper bound for the risk of $X + Y$ when risk is
measured by $\mathrm{ES}$. By comonotonic additivity, we know additionally that this
upper bound is sharp in the sense that it occurs in the case of comonotonic
$X$ and $Y$ (i.e. $X = f\circ Z$ and $Y = g\circ Z$ for some random variable
$Z$ and non-decreasing functions $f$ and $g$).

Compare this to the situation  when $\mathrm{VaR}_\alpha$ is used as risk
measure. Then there is no easy general upper bound for the risk of $X + Y$,
and finding the joint distribution of $X$ and $Y$ which yields the maximum
value for $\mathrm{VaR}_\alpha(X+Y)$ is a non-trivial task \citep{Embetal,
  LucMar}.
\end{remark}
Note that there are coherent and law invariant risk measures which are not
comonotonic additive (e.g. the standard semi-deviation, see \citealp{Fischer}).

It might have become clear from the above considerations that the class of
coherent, law invariant and comonotonic additive risk measures is of
particular interest. In \cite{kusuoka}, a complete characterization of this
class was accomplished, under the additional assumption that the risk measures
under consideration satisfy the so-called \emph{Fatou property}. We show that
this assumption is dispensable.
\begin{theorem}
  \label{th:1}
Let $\rho$ be a risk measure on the space $V$ of the bounded random variables
in the probability space $(\Omega, \mathcal{F}, \mathrm{P})$. Assume that
$(\Omega, \mathcal{F}, \mathrm{P})$ is standard and non-atomic (i.e.\ there
exists a random variable which is uniformly distributed on $(0,1)$). Then
$\rho$ is a coherent, law invariant and comonotonic additive (see Definition
\ref{de:10} and Proposition \ref{pr:10} (4), (5)) risk measure if and only if 
\begin{equation}
  \label{eq:44}
  \rho(X) = p \int_0^1 \mathrm{VaR}_u(X)\,F(d u) + (1-p)\,\mathrm{VaR}_1(X),\quad
  X\in V,
\end{equation}
where $p\in[0,1]$ and $F$ is a continuous convex distribution function which is concentrated on
$[0,1]$.
\end{theorem}
\begin{remark}\ 
  \label{rm:12} 
  \begin{enumerate}
  \item[(i)] Choose $p=1$ and $F(u) = \max\bigl( 0,
    \frac{u-\alpha}{1-\alpha}\bigr)$ in order to obtain $\mathrm{ES}_\alpha$
    from (\ref{eq:44}).
  \item[(ii)] Note that any continuous and convex distribution function $F$ on
    $[0,1]$ is absolutely continuous, i.e.\ can be written as $F(u) = \int_0^u
    f(t)\,d t$ where $f$ is its density with respect to Lebesgue measure. Thus
    Theorem \ref{th:1} states that the class of \emph{spectral risk measures}
    which was introduced in \cite{acerbi} is just the class of coherent, law invariant and
    comonotonic additive risk measures.
  \item[(iii)] Formulas like (\ref{eq:44}) can be traced back a long time in
    the actuarial literature \citep[cf.][and the references therein]{Wang}.
  \end{enumerate}
\end{remark}
\noindent\textbf{Proof of Theorem \ref{th:1}.} Let us first regard the case where a
risk measure $\rho$ as in (\ref{eq:44}) is given. Law invariance of $\rho$ is
then clear since (\ref{eq:44}) is based on quantiles of $X$.
If $p =0$ then $\rho$ is just the essential supremum of $X$. It is then
obvious that $\rho$ is coherent and comonotonic additive.

Assume now $p > 0$. Construct a function $F_0 : [0,1]\to[0,1]$ by setting
\begin{equation}
  \label{eq:45}
  F_0(u) = 
  \begin{cases}
    p\,F(u), & 0 \le u < 1\\
    1, & u = 1.
  \end{cases}
\end{equation}
Observe that $F_0$ is again convex and non-decreasing but may fail to be
continuous in 1. Nevertheless, 
it is easy to show that (\ref{eq:44}) is equivalent to
\begin{equation}
  \label{eq:46}
  \rho(X) = - \int X \,d F_0 \circ \mathrm{P}, \quad X\in V,
\end{equation}
where  $\int X \,d F_0 \circ \mathrm{P}$ denotes the non-additive integral
with respect to the \emph{distorted probability} $F_0 \circ \mathrm{P}$ in the
sense of \cite{De}. Coherence and comonotonic additivity of $\rho$ are now
just conclusions from the general theory of non-additive integration.

Next, we show that any coherent, law invariant and comonotonic additive risk
measure $\rho$ can be represented as in (\ref{eq:44}). As a first step, we
conclude from the results in \citet[][Schmeidler's theorem]{Schm86} that $\rho$
can be written as 
\begin{equation}
  \label{eq:47}
 \rho(X) = - \int X \,d\, \nu, \quad X\in V, 
\end{equation}
where $\int X \,d\, \nu$ denotes again a non-additive integral in the sense of
\citet{De}. $\nu$ is a monotonous (i.e.\ $A,B \in \mathcal{F}, A \subset B$
$\Rightarrow$ $\nu(A) \le \nu(B)$) and super-modular  (i.e.\ $A,B \in
\mathcal{F}$ $\Rightarrow$ $\nu(A) + \nu(B) \le \nu(A\cup B) + \nu(A \cap B)$)
set function (i.e.\ $\nu(\emptyset) = 0$) on $(\Omega, \mathcal{F})$ with
$\nu(\Omega) =1$. We define a function $F_0 : [0,1]\to[0,1]$ by
\begin{subequations}
  \begin{equation}
    \label{eq:48}
    F_0(u)\ = \ \nu(A)
  \end{equation}
for any $A\in\mathcal{F}$ with $\mathrm{P}[A] = u$. Since $(\Omega,
\mathcal{F}, \mathrm{P})$ is standard and non-atomic, for every $u\in [0,1]$
there is at least one $A\in\mathcal{F}$ with $\mathrm{P}[A] = u$. The law
invariance of $\rho$ implies that $F_0$ is with (\ref{eq:48}) is
well-defined. The monotonicity of $\nu$ implies that $F_0$ is
non-decreasing. Moreover, from (\ref{eq:48}) also follows
\begin{equation}
  \label{eq:48a}
  \nu = F_0 \circ \mathrm{P}.
\end{equation}
\end{subequations}
Again, since $(\Omega,
\mathcal{F}, \mathrm{P})$ is standard and non-atomic, for any $u_1,
u_2, u_3, u_4$ $\in [0,1]$ 
with $0\le u_1 < u_4 \le 1$, $u_2, u_3 \in[u_1,u_4]$
and $u_2 - u_1 = u_4 - u_3$ there are events $A,B \in \mathcal{F}$ such that
we have 
\begin{equation}
  \label{eq:49}
\mathrm{P}[A] = u_3,\ \mathrm{P}[B] = u_2, \  \mathrm{P}[A\cap B] = u_1,\
\mathrm{P}[A\cup B] = u_4. 
\end{equation}
The super-modularity of $\nu$ and (\ref{eq:49}) imply 
\begin{equation}
  \label{eq:50}
  F_0(u_4) + F_0(u_1) \ge F_0(u_2) + F_0(u_3)
\end{equation}
for all $u_1, u_2, u_3, u_4$ as above. With $u_2 = u_3$, (\ref{eq:50}) yields
for any $0 \le u < v \le 1$ that
\begin{subequations}
  \begin{equation}
    \label{eq:51}
F_0\bigl(\frac{u+v}2\bigr) \le \frac 1 2\,F_0(u) + \frac 1 2\,F_0(v).    
  \end{equation}
Of course, from  (\ref{eq:51})  we obtain 
\begin{equation}
  \label{eq:51a}
F_0\bigl(\alpha\,u+(1-\alpha)\,v\bigr) \le \alpha\,F_0(u) + (1-\alpha)\,F_0(v)  
\end{equation}
\end{subequations}
for every $\alpha \in\bigl\{\frac k {2^n} : n \ge 1, k = 0, 1, \ldots,
2^n\bigr\}$. Since $F_0$ is non-decreasing, limits from the right exist in
$\alpha\,u+(1-\alpha)\,v$
for every $\alpha \in (0,1)$. Hence, by passing to the limits in
(\ref{eq:51a}) we can conclude that (\ref{eq:51a}) holds for every $\alpha \in
[0,1]$, i.e.\ $F_0$ is convex. Observe that a function $F_0 : [0,1]\to[0,1]$
with $F_0(0) = 0$ and $F_0(1) = 1$ is necessarily continuous on $[0,1)$  if it
is non-decreasing and convex. Furthermore, $F_0$ can be constant at most on an
interval $[0,\epsilon)$. On $[\epsilon, 1]$ it will then be strictly
increasing.

So far, we know that $\rho$ can be represented by (\ref{eq:46}) where $F_0 :
[0,1]\to[0,1]$ is non-decreasing and convex as described above. Now, applying
the definition and some other properties of non-additive integrals yields
representation (\ref{eq:44}) where $p$ is given by $p = \sup_{u\in [0,1)}
F_0(u)$ and $F$ and $F_0$ are related by (\ref{eq:45}). \hfill $\Box$
\begin{remark}[Incorporating moment effects in ES]
  \label{rm:13} \ \\
Representation (\ref{eq:46}) allows in some cases a helpful
  interpretation of what happens when $\rho(X)$ is calculated. 
Fix any positive integer $n$. Recall from Remark \ref{rm:12}~(i) the function
$F$ which generates ES when used in (\ref{eq:44}). Define
  \begin{equation}
    \label{eq:1n}
    M_n(u)\, =\, F(1 - \sqrt[n]{1-u}), \quad u \in [0,1].
  \end{equation}
If $X$ is a real random variable, interpreted as the profit and loss of a
financial asset, consider independent and identically
distributed $Y, Y_1, \ldots, Y_n$  
with $\mathrm{P}[-Y \le t]= 1 - \sqrt[n]{\mathrm{P}[-X > t]}$. Then
\begin{align}
  \begin{split}
  \mathrm{P}[-X \le t]& = \mathrm{P}\bigl[\min(-Y_1, \ldots, -Y_n)\le
  t\bigr], \label{eq:3n}\\[1ex]
F\bigl(\mathrm{P}[Y > t]\bigr) & = M_n\bigl(\mathrm{P}[X > t]\bigr), 
\end{split}
  \intertext{and hence}
\label{eq:2n}
   \int Y\, d\, F\circ \mathrm{P} & = \int X\, d\, M_n\circ
  \mathrm{P}.
\end{align}
By Theorem \ref{th:1}, hence $\mathrm{ES}_\alpha^{(n)}(X)
=\mathrm{ES}_\alpha(Y)$ is a spectral risk measure. 
Note that \citep[cf.][]{Delb98}
\begin{subequations}
\begin{align}\label{eq:4n}
  \mathrm{E}[\max(0,-Y)] < \infty & \ \Rightarrow\
  \mathrm{E}\bigl[(\max(0,-X))^n\bigr] < \infty\\
\intertext{and for any $\epsilon > 0$}%
\mathrm{E}\bigl[(\max(0,-X))^{n+\epsilon}\bigr] < \infty & \ \Rightarrow\
\mathrm{E}[\max(0,-Y)] < \infty.  \label{eq:5n}
\end{align}
(\ref{eq:4n}) and (\ref{eq:5n}) show that $\mathrm{ES}_\alpha^{(n)}(X)$ is sensible to the $n$-th
moment of $X$. By (\ref{eq:3n}) it may be interpreted
as the Expected Shortfall
of a random variable $Y$ which is generated from $X$ by a pessimistic
manipulation since the loss variable $-X$ has the same distribution as the
minimum of $n$ independent copies of the loss $-Y$.
\end{subequations}
\end{remark}

\section{Risk contributions}
\label{sec:contr}

In this section we study the following problem: Given random variables $X_1,
\ldots, X_d$ (e.g.\ profits and losses of the different business lines in a
financial institution), portfolio weights $u_1, \ldots, u_d$, and a risk
measure $\rho$, we want to know how much $u_i\,X_i$ contributes to the total
risk $\rho\bigl( \sum_{i=1}^d u_i\,X_i\bigr)$ of the portfolio. With $u =
(u_1, \ldots, u_d)$ write for short 
\begin{equation}
  \label{eq:53}
\rho(u) =  \rho\bigl( \sum_{i=1}^d u_i\,X_i\bigr). 
\end{equation}
\cite{Denault} and \cite{T99} (with different reasonings) argued that
$u_i\,\frac{\partial \rho}{\partial u_i}(u)$ is the appropriate definition for
the \emph{risk contribution} of $u_i\,X_i$ in the case when $\rho(u)$ is
partially differentiable with respect to the components of $u$. 

The question of how to compute the partial derivatives in case $\rho =
\mathrm{VaR}_\alpha$ was independently tackled by several authors
\citep{GLS99, H99, L99, T99}. They observed that under certain smoothness
assumptions on the joint distribution of $(X_1,\ldots, X_d)$ 
\begin{subequations}
  \begin{equation}
    \label{eq:54}
\frac{\partial \mathrm{VaR}_\alpha}{\partial u_i}\Big( \sum_{j=1}^d
u_j\,X_j\Big) = -\,\mathrm{E}\Big[X_i\,\Big|\, - \sum_{j=1}^d
u_j\,X_j=  \mathrm{VaR}_\alpha  \bigl( \sum_{j=1}^d
u_j\,X_j\bigr)\Big].
  \end{equation}
Note that in case of integrable components $X_1,\ldots, X_d$ the right-hand
side of (\ref{eq:54}) will always be defined. By the factorization lemma there
are functions $\phi_i = \phi_i(u; z)$, $i=1, \ldots, d$, such that almost
surely
\begin{equation}
  \label{eq:54a}
\mathrm{E}\Big[X_i\,\Big|\, \sum_{j=1}^d
u_j\,X_j= z\Big] = \phi_i(u; z). 
\end{equation}
\end{subequations}
By inspection of (\ref{eq:44}), (\ref{eq:54}) and (\ref{eq:54a}) suggest the
following definition for risk contributions in case of a spectral risk measure
in the sense of Remark \ref{rm:12} (ii).
\begin{definition}[Risk contributions of spectral risk measures]
  \label{de:12}\sloppy \ \\
Let $X_1,\ldots, X_d$ be integrable random variables on the same probability
space, $u = (u_1, \ldots, u_d)$ their portfolio weight vector, and $\rho$ be a
spectral risk measure with representation (\ref{eq:44}). Define functions 
$\phi_i = \phi_i(u; z)$, $i=1, \ldots, d$, by (\ref{eq:54a}) and write
$\mathrm{VaR}_\alpha(u)$ for $\mathrm{VaR}_\alpha\bigl(\sum_{j=1}^d
u_j\,X_j\bigr)$.
Then, if all the involved integrals exist, the quantity
\begin{equation}
  \label{eq:55}
  \rho_i(u) = - p \int_0^1 \phi_i\bigl(u; - \mathrm{VaR}_\alpha(u)\bigr)\,F(d
  \alpha) - (1-p)\,\phi_i\bigl(u; - \mathrm{VaR}_1(u)\bigr)
\end{equation}
is called \emph{marginal impact} of $X_i$ on the total risk $\rho(u) =  
\rho\bigl( \sum_{j=1}^d u_j\,X_j\bigr)$. The quantity $u_i\,\rho_i(u)$ is
called \emph{risk contribution} of $u_i\,X_i$ to the total risk.
\end{definition}
By the standard theory of non-additive integration \citep[see][]{De} we obtain
the following equivalent representation of the risk contributions from
Definition \ref{de:12}.
\begin{proposition}
  \label{pr:14}
Assume that the random variables $X_1,\ldots, X_d$ from Definition \ref{de:12}
are defined on the probability space $(\Omega, \mathcal{F}, \mathrm{P})$. Then
the marginal impacts $\rho_i(u)$ in (\ref{eq:55}) can be equivalently written
as
\begin{equation}
  \label{eq:56}
\rho_i(u) = - \int \phi_i\bigl(u;\sum_{j=1}^d u_j\,X_j\bigr)\, d F_0 \circ \mathrm{P},
\end{equation}
where $F_0$ is given by (\ref{eq:45}). As a consequence, for fixed $u$, the value
of $\rho_i(u)$ does not depend on the choice of $\phi_i(u;\cdot)$.
\end{proposition}
\noindent\textbf{Proof.} (\ref{eq:55}) $\iff$ (\ref{eq:56}) can be proved like
in the proof of Theorem \ref{th:1}. Denote $\sum_{j=1}^d u_j\,X_j$ by $Z(u)$,
and let $\phi_i(u;z)$ and $\phi_i^\ast(u;z)$ be two versions of
$\mathrm{E}\bigl[X_i\,|\,Z(u) = z\bigr]$. Then,  $\phi_i(u;z)$ and
$\phi_i^\ast(u;z)$ are equal for all $z$ but those in a set $N$ with
$\mathrm{P}\bigl[Z(u) \in N \bigr] = 0$. Of course, then we also have $(F_0
\circ \mathrm{P})\bigl[Z(u) \in N \bigr] = 0$. This implies 
\begin{equation*}
\int \phi_i(u;Z(u)) \, d  F_0 \circ \mathrm{P} = \int \phi_i^\ast(u;Z(u)) \, d  F_0 \circ \mathrm{P}.
\end{equation*}
Thus, the proof is accomplished. \hfill $\Box$
\begin{remark}
  \label{rm:14}
(\ref{eq:56}) suggests the following procedure for the estimation of the
marginal impacts $\rho_i(u)$ on the spectral risk measure $\rho(u) =  
\rho\bigl( \sum_{j=1}^d u_j\,X_j\bigr)$:
\begin{enumerate}
\sloppy
\item Estimate the conditional expectations $\phi_i(u;\cdot)$ (see
  (\ref{eq:54a});
could be done by a kernel estimation).
\item Estimate the distribution of $\sum_{j=1}^d u_j\,X_j$ (could be done by a
  kernel estimation of the density).
\item Resample from the distribution of $\sum_{j=1}^d u_j\,X_j$ distorted by
  $F_0$, apply $\phi_i(u;\cdot)$ on the sample, and estimate $\rho_i(u)$ with
  the ordinary sample mean.
\end{enumerate}
\end{remark}
The representations (\ref{eq:55}) and (\ref{eq:56}) of the marginal impacts on
spectral risk measures can be significantly simplified in case of the Expected
Shortfall $\mathrm{ES}_\alpha$ (Definition \ref{de:11}). 
To see this we need the following two results.
\begin{proposition}
 \label{pr:1}
Let $X$ be a real random variable, $f: \mathbb{R} \to [0,\infty)$ a function
such that $\mathrm{E}\bigl[\max(0, - f\circ X)\bigr] < \infty$ and let
$\alpha \in (0,1)$ be a fixed confidence level. Then
\begin{multline}
  \label{eq:1}
\int_0^\alpha f\bigl(q_u(X)\bigr)\,d\, u\ =\ 
\mathrm{E}\bigl[ f\circ X \,\mathbf{1}_{\{X \le q_\alpha(X)\}}\bigr]\\ +
f\bigl(q_\alpha(X)\bigr)\,\bigl(\alpha - \mathrm{P}[X \le q_\alpha(X)]\bigr).
\end{multline}
\end{proposition}
\noindent \textbf{Proof.} By switching to another probability space if necessary, we can 
assume that there is a real random variable $U$  that
is uniformly distributed on $(0,1)$, i.e. $\mathrm{P}[U \le u] = u$, $u \in (0,1)$. 
It is well-known that then the random
variable $Z = q_U(X)$ has the same distribution as $X$.

Since $u \mapsto q_u(X) $ is non-decreasing we have
  \begin{equation}
  \label{eq:3}
  \begin{split}
\{ U \le \alpha\} & \subset  \{ Z \le q_\alpha(X)\}\quad \mbox{and}\\
\{ U >  \alpha\} \cap \{ Z \le q_\alpha(X)\} & \subset 
\{ Z = q_\alpha(X)\}\,.    
  \end{split}
\end{equation}
By (\ref{eq:3}) we obtain
\begin{equation}
\label{eq:5}
  \begin{split}
\int_0^\alpha f\bigl(q_u(X)\bigr)\,d\, u & =  \mathrm{E}[f\circ Z\,\mathbf{1}_{\{U \le \alpha\}}]\\
& =  \mathrm{E}[f\circ Z\,\mathbf{1}_{\{Z \le q_\alpha(X)\}}] 
- \mathrm{E}[f\circ Z\,\mathbf{1}_{\{ U >  \alpha\} \cap\{Z \le q_\alpha(X)\}}] \\
& =  \mathrm{E}[f\circ X\,\mathbf{1}_{\{X \le q_\alpha(X)\}}] + 
q_\alpha(X)\,\bigl( \alpha - \mathrm{P}[X \le q_\alpha(X)]\bigr).
    \end{split}
\end{equation}
Thus, the proof is accomplished. \hfill $\Box$
\begin{remark}
  \label{rm:1} 
Prop.\ \ref{pr:1} generalizes Prop.\ 3.2 of \citet{AT01}.
The ``$\le$'' in (\ref{eq:1})  may be replaced by ``$<$''.
\end{remark}
\begin{corollary}
  \label{co:1}
Let $X,Y$ be real random variables such that $\mathrm{E}\bigl[|Y|\bigr] <
\infty$ and let $\alpha \in (0,1)$ be a fixed confidence level. Then
\begin{multline}
  \label{eq:60}
\int_\alpha^1 \mathrm{E}\bigl[ Y\,|\,X = - \mathrm{VaR}_u(X)\bigr]\, d u = 
\mathrm{E}\bigl[Y\,\mathbf{1}_{\{- X \ge q_\alpha(- X)\}}\bigr] \\
+ \mathrm{E}\bigl[ Y\,|\,- X = q_\alpha(- X) \bigr]\Big(\mathrm{P}\bigl[ - X <
q_\alpha(- X)\bigr] -\alpha\Big).
\end{multline}
Moreover, the value of $\int_\alpha^1 \mathrm{E}\bigl[ Y\,|\,X = -
\mathrm{VaR}_u(X)\bigr]\, d u$
is the same for any version of the conditional expectation.
\end{corollary}
\noindent\textbf{Proof.} Non-dependence on the particular version of
conditional expectation follows from Proposition \ref{pr:1}. Observe that
\begin{subequations}
\begin{equation}
  \label{eq:61}
  \begin{split}
\int_\alpha^1 \mathrm{E}\bigl[ Y\,|\,X = - \mathrm{VaR}_u(X)\bigr]\, d u & = 
\int_0^1  \mathrm{E}\bigl[ Y\,|\,- X = q_u(- X)\bigr]\, d u\\
& \qquad -  \int_0^\alpha  \mathrm{E}\bigl[ Y\,|\,- X = q_u(- X)\bigr]\, d u \\
& = \mathrm{E}[Y] -\int_0^\alpha  \mathrm{E}\bigl[ Y\,|\,- X = q_u(- X)\bigr]\, d u.
  \end{split}
\end{equation}
Proposition \ref{pr:1} and Remark \ref{rm:1}, applied to $f(x) =
\mathrm{E}\bigl[Y \,|\, - X = x\bigr]$, yield
\begin{multline}
\label{eq:61a} 
\int_0^\alpha  \mathrm{E}\bigl[ Y\,|\,- X = q_u(- X)\bigr]\, d u =
\mathrm{E}\bigl[Y\,\mathbf{1}_{\{- X < q_\alpha(- X)\}}\bigr] \\
+ \mathrm{E}\bigl[ Y\,|\,- X = q_\alpha(- X) \bigr]\Big(\alpha - \mathrm{P}\bigl[ - X <
q_\alpha(- X)\bigr]\Big). 
\end{multline}  
\end{subequations}
(\ref{eq:61}) and (\ref{eq:61a}) imply the assertion. \hfill $\Box$

Recall Definition \ref{de:12} of the marginal impact $\rho_i(u)$ of a
component $X_i$ on the total risk $\rho(u)$ of a portfolio $\sum_{j=1}^d
u_j\,X_j$ when $\rho$ is a spectral risk measure. This definition applies to
$\rho = \mathrm{ES}_\alpha$ with $p = 1$ and $F(u) = \max\bigl( 0,
    \frac{u-\alpha}{1-\alpha}\bigr)$. In this case, Corollary \ref{co:1}
    implies the representation (with $Z(u) = \sum_{j=1}^d u_j\,X_j$) 
    \begin{multline}
     \label{eq:62}
\rho_i(u) = - (1 -\alpha)^{-1} \bigg\{ 
\mathrm{E}\bigl[X_i\,\mathbf{1}_{\{- Z(u) \ge q_\alpha(- Z(u))\}}\bigr] \\
+ \mathrm{E}\bigl[ X_i\,|\,- Z(u) = q_\alpha(- Z(u)) \bigr]\Big(\mathrm{P}\bigl[ - Z(u) <
q_\alpha(- Z(u))\bigr] -\alpha\Big)\bigg\}.
    \end{multline}
Returning to (\ref{eq:60}), we will show that its right-hand side times $(1
-\alpha)^{-1}$ (and, as a consequence, also $\rho_i(u)$ from(\ref{eq:62})) can
be interpreted as a conditional expectation given that a certain worst case
event has occurred. Note first that by the very definition of quantile we have
\begin{subequations}
  \begin{equation}
    \label{eq:63}
    0 \le \mathrm{P}\bigl[- X \le q_\alpha (- X)\bigr] - \alpha \le 
\mathrm{P}\bigl[- X =   q_\alpha (- X)\bigr]
  \end{equation}
and in particular
\begin{equation}
  \label{eq:63a}
 \mathrm{P}\bigl[- X \le q_\alpha (- X)\bigr] - \alpha \not= 0 \ \Rightarrow \
\mathrm{P}\bigl[- X =   q_\alpha (- X)\bigr] > 0.
\end{equation}
\end{subequations}
Hence, it makes sense to define a $\{0,1\}$-valued random variable $J = J_{X,\alpha}$ with
\begin{subequations}
  \begin{equation}
    \label{eq:16}
\mathrm{P}[J = 1] \ =\ p_\alpha \ = \ 1 - \mathrm{P}[J = 0],
  \end{equation}
where
\begin{equation}
  \label{eq:17}
p_\alpha \ = \ 
\begin{cases}
  \frac{\mathrm{P}[- X \le q_\alpha(- X)]-\alpha}{\mathrm{P}[- X = q_\alpha(-
    X)]}, & 
\text{if $\mathrm{P}[- X = q_\alpha(- X)] > 0$\,,}\\
0\,, & \text{otherwise.} 
\end{cases}
\end{equation}
\end{subequations}
\begin{proposition}
  \label{pr:2}
Let $X, Y$ be real random variables such that $\mathrm{E}[|Y|] < \infty$ and 
$\alpha \in (0,1)$ a fixed confidence level. Suppose that there is a random variable
$J$ which satisfies (\ref{eq:16}), (\ref{eq:17}) and is independent from $(X,Y)$.
Define 
\begin{subequations}
\begin{equation}
  \label{eq:18}
I \ =\ \mathbf{1}_{\{- X > q_\alpha(- X)\}\cup\{- X = q_\alpha(- X), J = 1\}}.
\end{equation}
Then
\begin{equation}
  \label{eq:19}
\mathrm{E}[Y\,|\,I = 1] \ =\ (1-\alpha)^{-1}\int_\alpha^1 
\mathrm{E}\bigl[Y\,|\, X = - \mathrm{VaR}_u(X)\bigr]\, d\, u.
\end{equation}
\end{subequations}
\end{proposition}
\noindent \textbf{Proof.} By (\ref{eq:17}) and the independence of $I$ and $(X,Y)$ we have
\begin{equation}
  \label{eq:20}
\mathrm{P}[I=1] \ =\ 1-\alpha.
\end{equation}
It is now straightforward to see that
\begin{multline}
  \label{eq:21}
\mathrm{E}[Y\,\mathbf{1}_{\{I = 1\}}]  =
\mathrm{E}[Y\,\mathbf{1}_{\{- X \ge q_\alpha(- X)\}}]\\
+ 
\mathrm{E}\bigl[Y\,|\,- X=q_\alpha(- X)\bigr]\,\Big(\mathrm{P}[- X <
q_\alpha(- X)] - \alpha\Big).
\end{multline}
Thus, the assertion follows from Corollary \ref{co:1}.\hfill $\Box$

The philosophy behind value-at-risk (VaR) is that the event $\{- X \le
q_\alpha(- X)\}$ is tolerable
whereas $\{- X > q_\alpha(- X)\}$ corresponds to a kind of default.
Note that
\begin{equation}
  \label{eq:22}
\mathrm{P}\bigl[- X > q_\alpha(- X)\bigr] \ \le \ 1-\alpha.
\end{equation}
Hence one might consider $I$ from Proposition \ref{pr:2} an indicator of $\{- X > q_\alpha(- X)\}$ 
modified in a way that enlarges
the probability of default. Setting $Y=X$ in (\ref{eq:19}) shows that ES itself may be regarded as 
a conditional expectation in a worst case scenario. Replacing $X$ by $Z(u)$ and $Y$ by $X_i$ shows
that the same holds for the ES marginal impacts from (\ref{eq:62}).

Observe that (\ref{eq:19}) is also a statement about how to estimate ES and
the ES marginal impacts. Assume that an independent, identically distributed 
sample $(X_{1,i}, \ldots, X_{d,i})$, $i = 1, \ldots, N$,
of the portfolio component returns is given (cf. (\ref{eq:62})). 
Let $Z_i = \sum_{j=1}^d u_j\,X_{j,i}$, $i = 1, \ldots, N$.
\begin{itemize}
\item First estimate $q_\alpha(- Z)$ from $(Z_1, \ldots, Z_N)$ by some number $\hat{q}$. 
\item Estimate the probabilities $\mathrm{P}\bigl[- Z \le q_\alpha(- Z)\bigr]$
  and 
$\mathrm{P}[- Z = q_\alpha(- Z)]$.
Let $p_{s}$ and $p_e$ denote the corresponding estimators.
\item Determine a sub-sample by taking all those $i$ such that $- Z_i >
  \hat{q}$ or $- Z_i = \hat{q}$ and
an additional independent Bernoulli experiment with success probability
$\frac{p_s -\alpha}{p_e}$
(only in case $p_e > 0$) results in 1.
\item Estimate $\mathrm{ES}_\alpha(Z)$ and the marginal impacts according to
  (\ref{eq:62}) 
as negative averages of this
sub-sample.
\end{itemize}


\end{document}